\documentclass[twocolumn,superscriptaddress,nofootinbib,floatfix,aps,prb,citeautoscript,longbibliography]{revtex4-2}
\usepackage[utf8]{inputenc}
\usepackage[T1]{fontenc}
\usepackage{lineno}

\usepackage[colorlinks=true,linkcolor=blue,citecolor=blue,urlcolor=blue]{hyperref}
\usepackage{graphicx}%[draft]
\usepackage{xcolor}
\usepackage{xstring}
\usepackage{array}
\usepackage{dcolumn}
\usepackage{booktabs}
\usepackage{bm}
\usepackage{multirow}
\usepackage[version=4]{mhchem}
\usepackage{cleveref}
\usepackage{orcidlink}
\usepackage{float}

\newcommand{\Tc}{\ensuremath{T_\textrm{c}}}

\newcommand{\olog}{\ensuremath{\omega_\text{log}}}

\newcommand{\la}{\ensuremath{\lambda}}

\newcommand{\SIagm}[1]{%
\IfEqCase{#1}{%
{agm006249793}{SI \#1}%
{agm006249797}{SI \#2}%
{agm006249802}{SI \#3}%
{agm006249795}{SI \#4}%
{agm073026977}{SI \#5}%
{agm073024595}{SI \#6}%
{agm006249791}{SI \#7}%
{agm003245870}{SI \#8}%
{agm073025883}{SI \#9}%
{agm006249799}{SI \#10}%
{agm006249775}{SI \#11}%
{agm006249778}{SI \#12}%
{agm085198323}{SI \#13}%
{agm085198331}{SI \#14}%
{agm085198344}{SI \#15}%
}[AGM NOT FOUND]}

\begin{document}

\newcommand{\bochum}{Research Center Future Energy Materials and Systems of the University Alliance Ruhr and Interdisciplinary Centre for Advanced Materials Simulation, Ruhr University Bochum, Universit\"atsstraße 150, D-44801 Bochum, Germany}
\newcommand{\coimbra}{CFisUC, Department of Physics, University of Coimbra, Rua Larga, 3004-516 Coimbra, Portugal}
\newcommand{\mpihalle}{Max-Planck-Institut f\"ur Mikrostrukturphysik, Weinberg 2, D-06120 Halle, Germany}
\newcommand{\basque}{Fisika Aplikatua Saila, University of the Basque Country (UPV/EHU), 20018 Donostia/San Sebasti\'an, Spain}
\newcommand{\jsnu}{Laboratory of Quantum Materials Design and Application, School of Physics and Electronic Engineering, Jiangsu Normal University, Xuzhou 221116, China}

\author{Kun Gao\orcidlink{0009-0007-6032-9683}}
\affiliation{\bochum} 
\author{Wenwen Cui\orcidlink{0000-0003-4097-1146}}
\email{wenwencui@jsnu.edu.cn}
\affiliation{\jsnu} 
\author{Tiago F. T. Cerqueira\orcidlink{0000-0002-4147-8129}}
\affiliation{\coimbra}
\author{Hai-Chen Wang\orcidlink{0000-0002-2892-5879}}
\affiliation{\bochum}
\author{Silvana Botti\orcidlink{0000-0002-4920-2370}}
\affiliation{\bochum} 
\author{Miguel A. L. Marques\orcidlink{0000-0003-0170-8222}} 
\email{miguel.marques@rub.de}
\affiliation{\bochum} 

\title{Enhanced superconductivity in X$_4$H$_{15}$ compounds via hole-doping at ambient pressure }

\begin{abstract}
This study presents a computational investigation of \ce{X4H15} compounds (where X represents a metal) as potential superconductors at ambient conditions or under pressure. Through systematic density functional theory calculations and electron-phonon coupling analysis, we demonstrate that electronic structure engineering via hole doping dramatically enhances the superconducting properties of these materials. While electron-doped compounds with X$^{4+}$ cations (Ti, Zr, Hf, Th) exhibit modest transition temperatures of 1--9 K, hole-doped systems with X$^{3+}$ cations (Y, Tb, Dy, Ho, Er, Tm, Lu) show remarkably higher values of approximately 50~K at ambient pressure. Superconductivity in hole-doped compounds originates from stronger coupling between electrons and both cation and hydrogen phonon modes. Although pristine \ce{X$^{3+}$4H15} compounds are thermodynamically unstable, we propose a viable synthesis route via controlled hole doping of the charge-compensated \ce{YZr3H15} compound. Our calculations predict that even minimal concentrations of excess Y could induce high-temperature superconductivity while preserving structural integrity. This work reveals how strategic electronic structure modulation can optimize superconducting properties in hydride systems, establishing a promising pathway toward practical high-temperature conventional superconductors at ambient pressure.
\end{abstract}

\maketitle

\section{Introduction}
Hydride superconductors have attracted significant attention due to their potential for achieving high-temperature superconductivity, particularly in high-pressure synthesized binary hydrides, such as H$_3$S (203~K, 155~GPa) ~\cite{li2014metallization,duan2014pressure,drozdov2015conventional} and LaH$_{10}$  (250-260~K, 170~GPa)~\cite{liu2017potential,peng2017hydrogen,drozdov2019superconductivity,somayazulu2019evidence}, and ternary hydrides, such as (La, Y)H$_{10}$ (253 K at 183 GPa) \cite{semenok2021superconductivity}, (La, Ca)H$_{10}$ (247 K at 173 GPa) \cite{chen2024synthesis}, and (La, Al)H$_{10}$ (223 K at 164 GPa) \cite{chen2024high}.
These structures exhibit remarkably high superconducting transition temperature (\Tc), yet their stabilization pressures exceed 150~GPa, posing a significant barrier to practical applications. Consequently, researchers have shifted their focus toward identifying hydrides that maintain stability under relatively low-pressure conditions or even at ambient temperatures, with the goal of discovering practical, novel superconductors.

Among these systems, the \ce{X4H15} compounds (where X = Zr, Hf, Th) represent a captivating class of materials, requiring relatively moderate pressures compared to other hydride superconductors. \ce{Th4H15}, first synthesized and characterized in the 1970s, exhibits a transition temperature (\Tc) of 7.5--8 K at ambient pressure, establishing it as a pioneering example of hydride superconductivity~\cite{satterthwaite1970superconductivity, satterthwaite1972preparation}. The Th sublattice in \ce{Th4H15} adopts a cI16 structure, characterized as a distorted $2\times2\times2$ supercell of a body-centered cubic (bcc) sublattice~\cite{satterthwaite1972preparation}. The hydrogen atoms in \ce{Th4H15} form an intricate network, incorporating both interstitial and framework hydrogen. This distinctive hydrogen arrangement, with multiple interstitial site occupations, profoundly influences the electronic and vibrational properties, thereby controlling superconductivity. Subsequently, \ce{Hf4H15} and \ce{Zr4H15} were successfully synthesized with analogous structures~\cite{kuzovnikov2019high,xie2020superconducting}, confirming the robustness and versatility of this hydride class. Experimental investigations reveal that \ce{Hf4H15} achieves a \Tc\ of 4.5~K at 23~GPa~\cite{kuzovnikov2019high}, while \ce{Zr4H15} demonstrates superconductivity with a \Tc\ of 4~K at 40~GPa~\cite{xie2020superconducting}. Computational studies have provided crucial insights into the stability and electronic properties of these systems. Theoretical calculations for \ce{Hf4H15} predict superconductivity with \Tc\ ranging from 0.8 to 2.1 K at 200 GPa, maintaining structural stability between 100-200 GPa~\cite{xie2020hydrogen}. Similarly, simulations for \ce{Zr4H15} predict a \Tc\ of 0.2-0.8~K at 40 GPa, with stability in the 50--100 GPa range~\cite{xie2020superconducting}.

The uniqueness of the \ce{X4H15} structure resides in its high hydrogen content and distinctive electronic properties, positioning it as an ideal candidate for investigating superconductivity in low-pressure or ambient-pressure hydrides. In this work, we present a comprehensive computational investigation of the \ce{X4H15} family, methodically exploring potential compounds where X spans the periodic table. Our study integrates thermodynamic stability analysis with electron-phonon coupling calculations to predict superconducting properties. The objective is to elucidate the fundamental mechanisms driving superconductivity and develop optimization strategies for achieving higher \Tc. Our investigation reveals that hole doping emerges as a remarkably effective approach for enhancing superconductivity. By introducing hole carriers, selective doping strategically modulates the electronic density of states (DOS) near the Fermi level and significantly strengthens electron-phonon coupling, offering a powerful new paradigm for optimizing the superconducting properties of hydrides \cite{PhysRevB.104.L020504, PhysRevB.109.014502,GE2020100330}.

We remark that previous theoretical studies have demonstrated that hole doping in hydrides such as \ce{CaYH12}~\cite{PhysRevB.99.100505}, \ce{Ca(BH4)2}~\cite{PhysRevB.107.L060501} can strongly increase \Tc. This enhancement primarily stems from critical modifications in the DOS near the Fermi level and intensification of phonon softening effects induced by hole doping, which synergistically amplify electron-phonon coupling. Furthermore, hole doping not only potentially triggers new superconducting phases but also substantially reduces the pressure threshold required for achieving high-temperature superconductivity, hence offering a viable pathway for exploring high-\Tc\ materials under ambient or near-ambient conditions \cite{ding2024hydrogen,chen2017enhanced}. Experimentally, significant advancements have been made in realizing hole doping through elemental substitution or the introduction of vacancy defects. For instance, in the \ce{LaH10} system, partial substitution of La with low-valence elements such as Be effectively introduces hole carriers, optimizing electron-phonon coupling and enhancing superconducting performance \cite{song2023stoichiometric}. Additionally, hole doping in hydrogen storage materials like \ce{Mg(BH4)2} has successfully induced an insulator-to-metal transition, accompanied by an exceptionally high DOS near the Fermi level, resulting in a \Tc\ as high as 140~K \cite{LIU2024101299}.

\section{Results and Discussion}

\begin{figure}[h] 
    \centering
        \begin{tabular}{c c}
        \includegraphics[width=0.45\columnwidth]{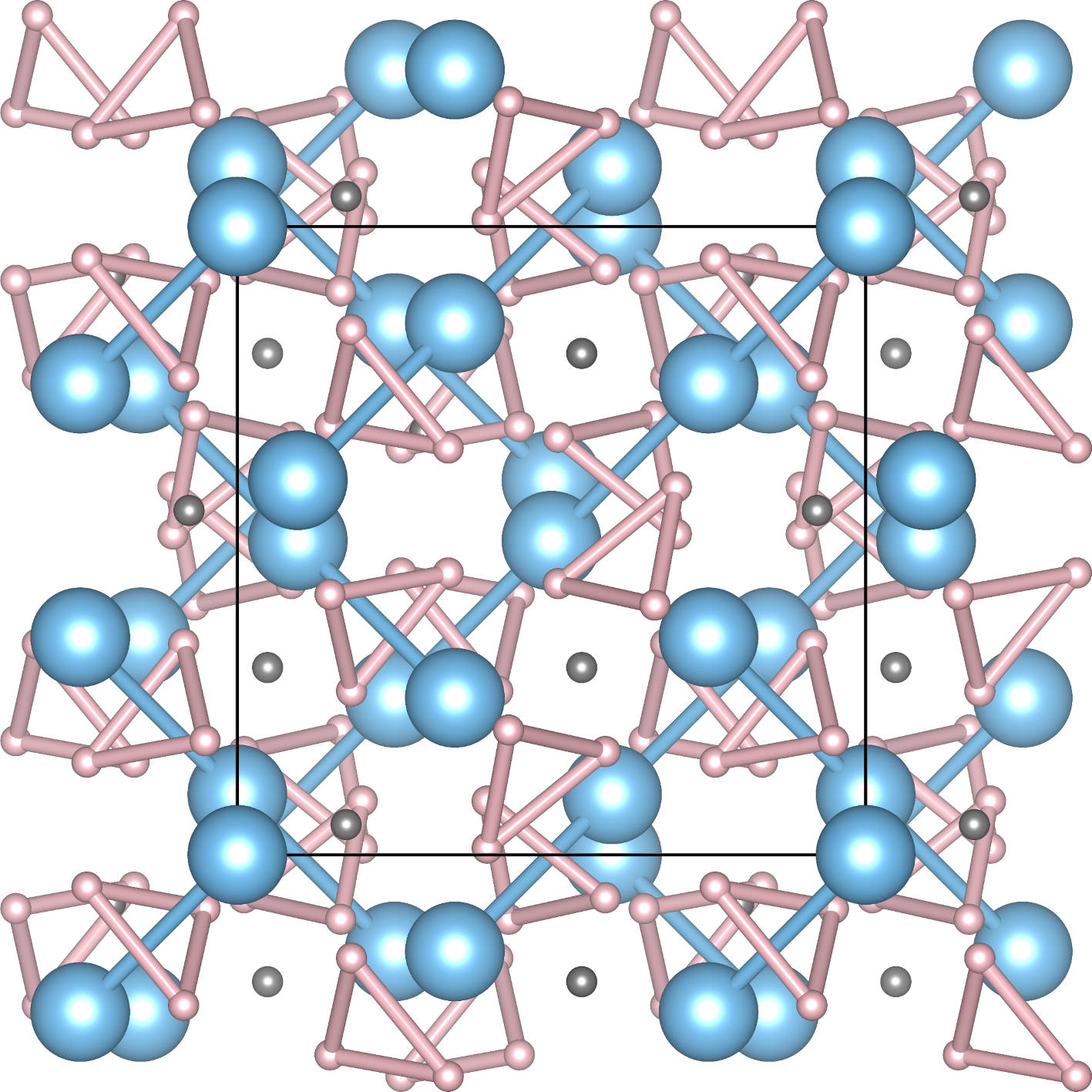} & \includegraphics[width=0.45\columnwidth]{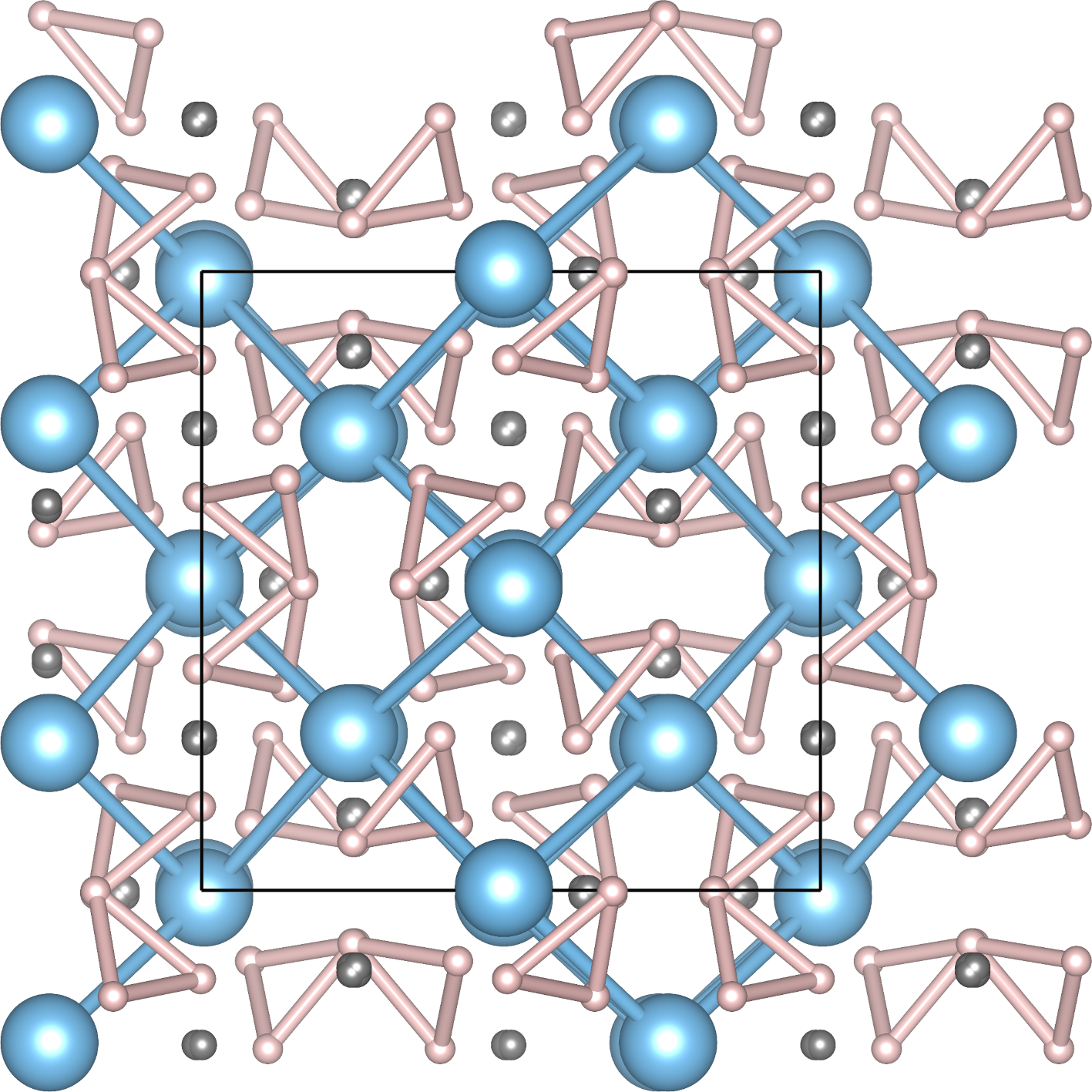}\\\vspace{-3mm}        
        (a) & (b) \\
        \end{tabular}
    \caption{ The structures of \ce{X4H15} with space group $I\bar43d$ at (a) 0~GPa and (b) 80~GPa. Blue, pink, and gray spheres denote the X atoms at 16c, H atoms at 48e, and H atoms at 12a Wyckoff positions, respectively. Note the bonds between hydrogen atoms are drawn only for eye guidance, the actual distance (>1.5~\AA) between them are much longer than the bond length in \ce{H2} molecule.}
    \label{fig:structure}
\end{figure}

The compound \ce{X4H15} crystallizes in the cubic system with space group $I\bar{4}3d$ (\#220) (see \cref{fig:structure}a). The X atoms exclusively occupy the 16$c$ Wyckoff positions with full occupancy, while the hydrogen atoms distribute across two distinct crystallographic sites: the general $48e$ positions and the more symmetric $12a$ positions. This arrangement yields a total of 16 metal X atoms and 60 H atoms per conventional unit cell (4 formula units). The distorted bcc configuration of the metal atoms, combined with the point group operations, generates an intricate three-dimensional network of X--H bonds. The strategic positioning of hydrogen atoms around the X centers creates well-defined channels for electron transport and phonon propagation, playing a decisive role in determining the superconducting properties of these materials.

Assuming the -1 oxidation state of hydrogen typical of hydrides, achieving charge neutrality with the \ce{X4H15} stoichiometry proves impossible, as pristine X$_4$H$_{15}$ would necessitate an oxidation state of +3.75 for the X atom. The three experimentally confirmed systems incorporate X atoms with oxidation state +4, resulting in one excess electron per formula unit relative to charge compensation. Conversely, a +3 cation generates three holes, while other oxidation states produce an untenable quantity of electrons or holes that would likely destabilize the system. Consequently, it is logical that all \ce{X4H15} systems exhibiting semiconducting behavior in our calculations (\ce{B4H15}, \ce{Cd4H15}, \ce{N4H15}, \ce{Tl4H15}, \ce{Bi4H15}, \ce{C4H15}, \ce{O4H15}, \ce{Si4H15}, \ce{Cu4H15}, \ce{Os4H15}, \ce{Ru4H15}, and \ce{La4H15}) relax to crystal structures that deviate substantially from the prototypical \ce{Th4H15} arrangement.

A number of \ce{X4H15} are magnetic, which hinders conventional superconductivity, specifically the compounds with X = K, Mn, Tc, Cr, V, Ce, Re, Bi, Gd, U, Pu, Np. Of the remaining 36 compounds, that are metallic and non-magnetic, only \ce{Zr4H15}, \ce{Hf4H15}, \ce{Th4H15}, and \ce{U4H15} are on the convex hull of stability. Note that the first three have been synthesized experimentally, in excellent agreement with our results. Up to 100~meV/atom of the convex hull we still find \ce{Np4H15} (26~meV/atom), \ce{Ce4H15} (37~meV/atom), \ce{Pu4H15} (81~meV/atom), \ce{K4H15} (91~meV/atom), and \ce{Ti4H15} (95~meV/atom). 

\begin{figure*}[htbp]
\centering
\begin{tabular}{c c c}
\includegraphics[width=0.33\textwidth]{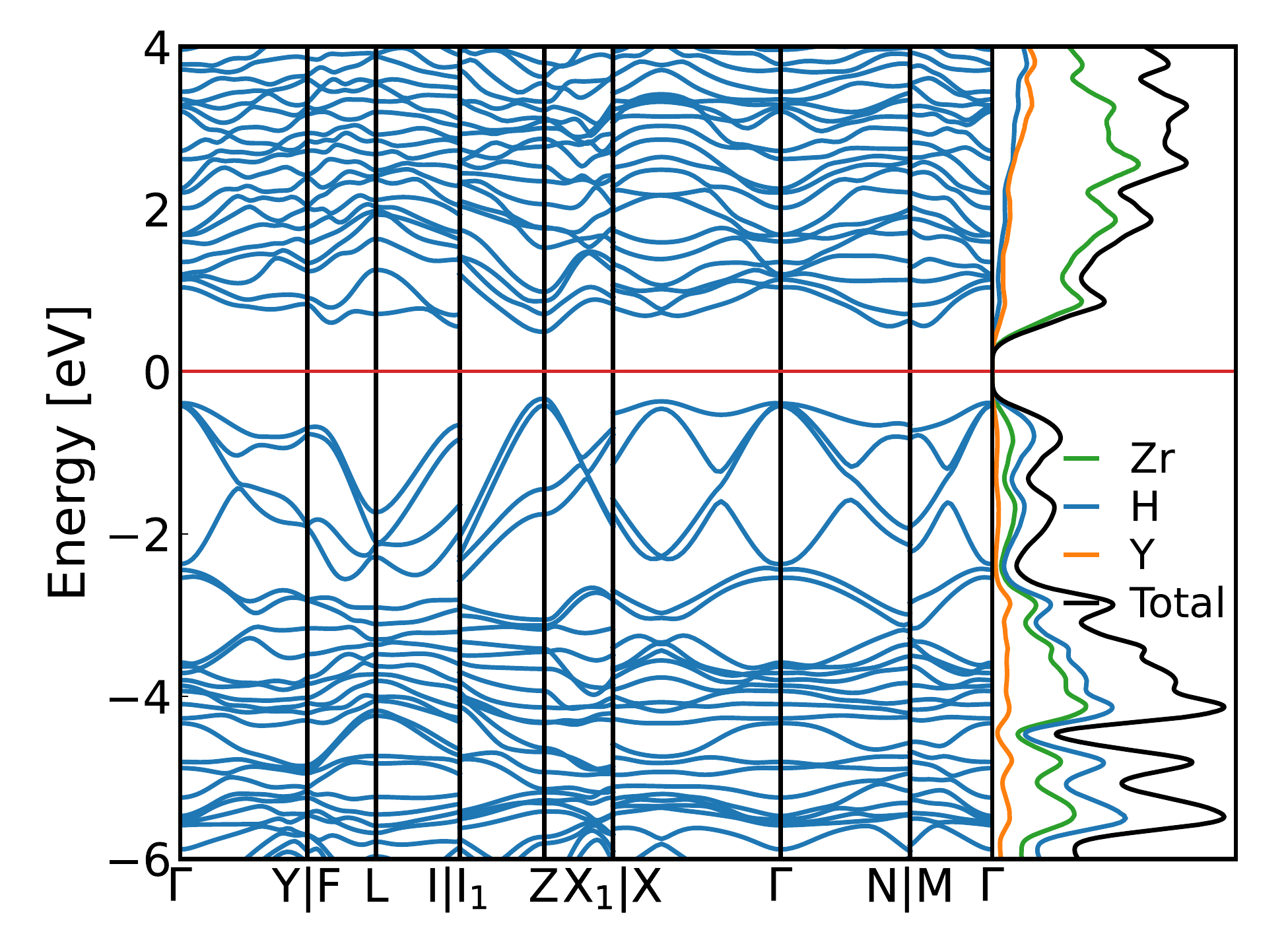}&
\includegraphics[width=0.33\textwidth]{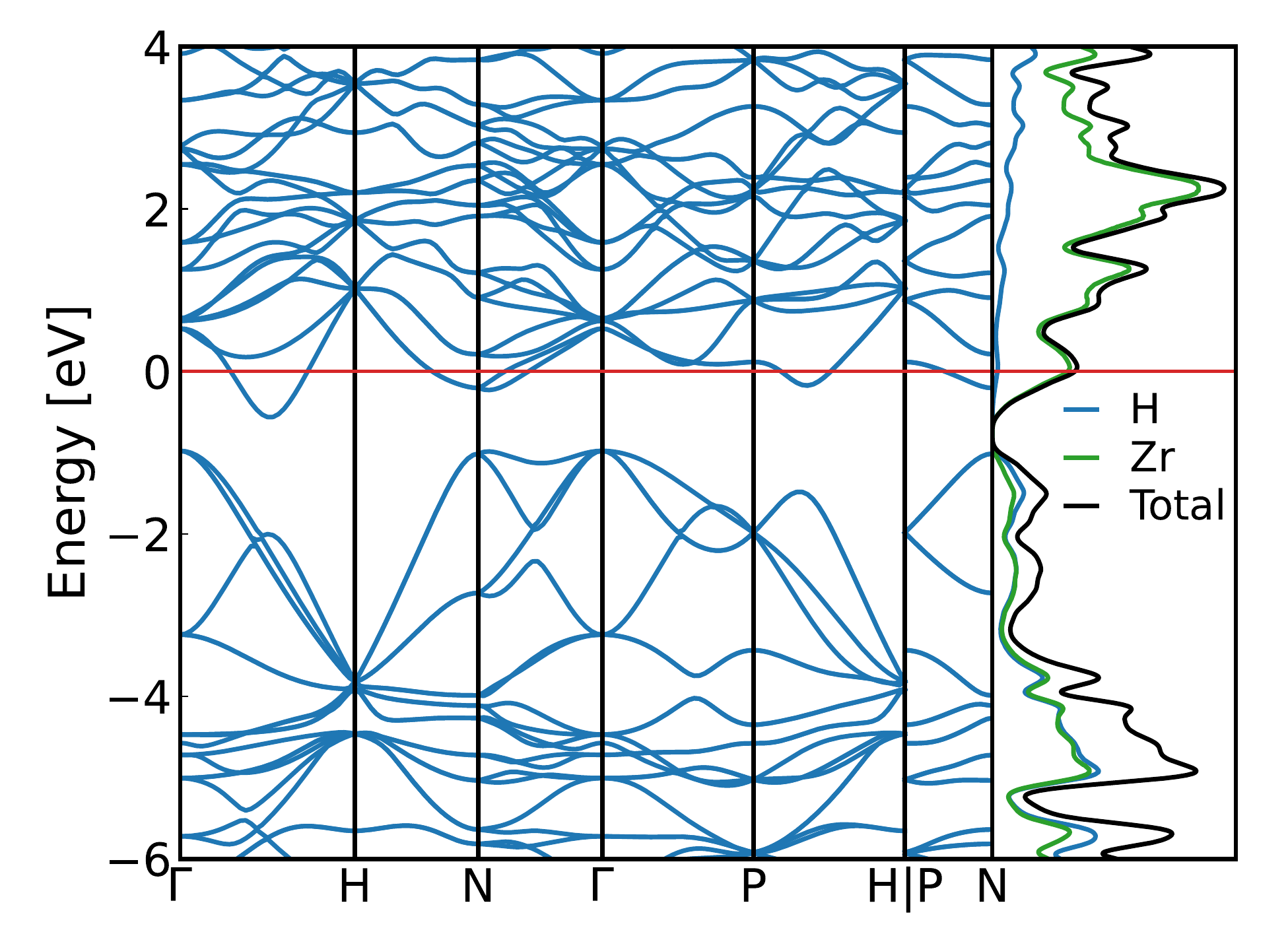}&
\includegraphics[width=0.33\textwidth]{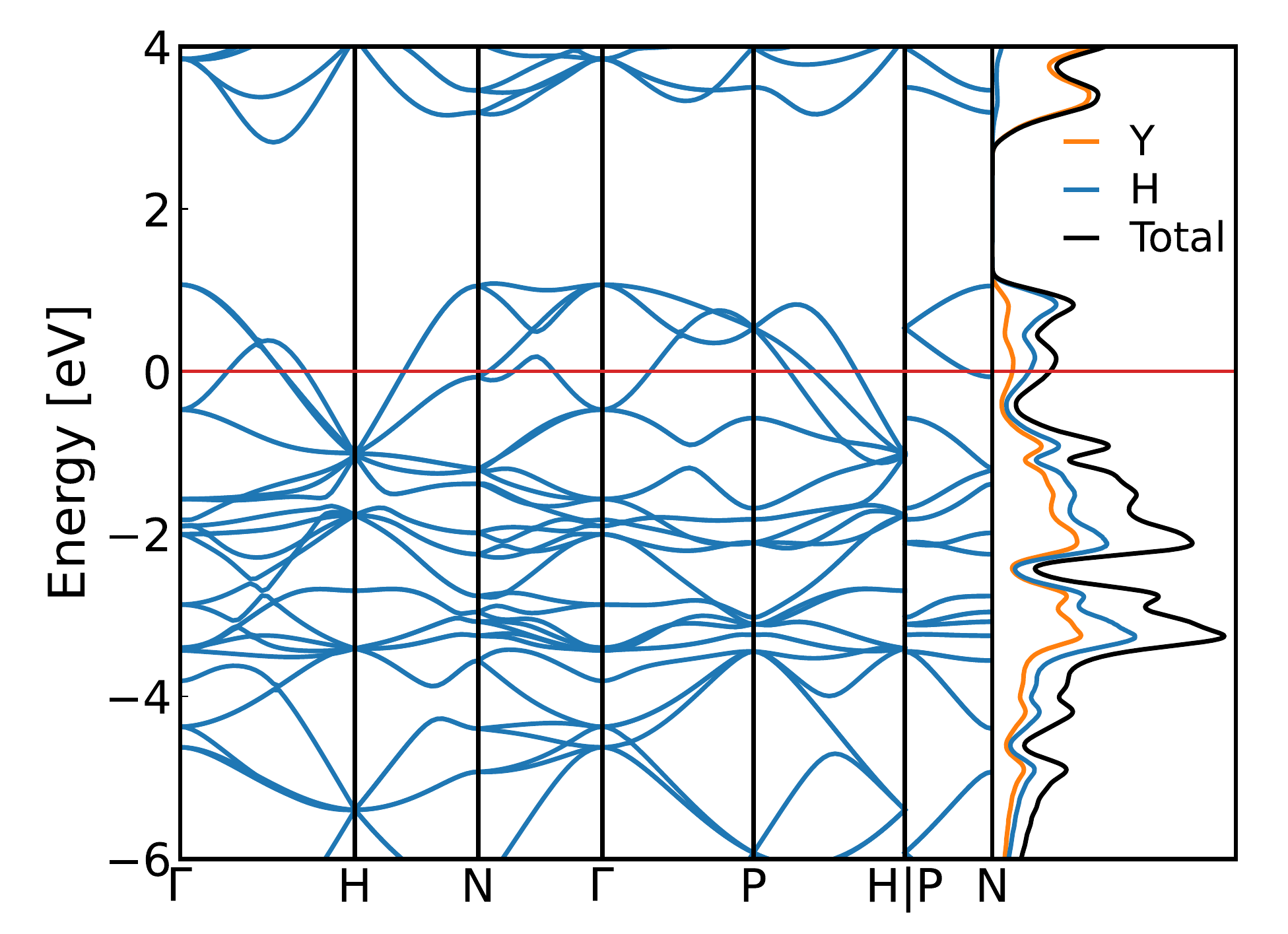} \\
(a)&(b)&(c)\\
\end{tabular}
\caption{Electronic band structure and density of states (DOS) for representative compounds: (a) charge-compensated \ce{YZr3H15}, (b) electron-doped \ce{Zr4H15}, and (c) hole-doped \ce{Y4H15}. The Fermi level is set at 0~eV.}
\label{fig:el-bs}
\end{figure*}

A possible way to improve the thermodynamic stability of this compounds is by enforcing charge compensation. To demonstrate this concept, we constructed YZr$_3$H$_{15}$ by substituting one Zr$^{4+}$ in Zr$_4$H$_{15}$ with Y$^{3+}$. This turns out to be a very stable compound, positioned merely 5~meV/atom from the convex hull of stability, even if \ce{Y4H15} is 233~meV/atom above the hull. As anticipated, this compound behaves as a semiconductor with an indirect band gap of approximately 1.1~eV within the PBE approximation (see \cref{fig:el-bs}a). The valence band exhibits triple degeneracy at $\Gamma$, with one band displaying large hole mass along specific directions in the Brillouin zone. This characteristic generates a steep increase in the electronic density of states immediately below the valence band maximum. These bands predominantly comprise hydrogen states hybridized with the cations. In contrast, the conduction bands manifest primarily Zr character, with minimal contributions from both Y and H. Based on these fundamental electronic structure characteristics, we can readily anticipate that hole doping would prove substantially more effective than electron doping in facilitating conventional superconductivity in these materials.

\begin{table*}
\caption{Summary of the \ce{X4H15} compounds that are dynamically stable at zero pressure. We show the oxidation state of the X cation, the distance above the convex hull of stability ($E_{\text{hull}}$ in eV/atom), the superconducting transition temperature calculated with the Allen-Dynes correction~\cite{AD} to the McMillan formula~\cite{McM} ($T_{c}^{\text{AD}}$ in K), the electron-phonon coupling constant $\lambda$, the logarithmic average of the phonon frequency ($\omega_{\text{log}}$ in K), the total density of electronic states are Fermi level (TDOS in states/eV/cell), and the partial density of H and X states ate the Fermi level (in states/eV/cell).}
\label{table:superconductors}
\begin{tabular*}{0.8\textwidth}{@{\extracolsep{\fill}}l cccccccccc}
\toprule
Compound &  oxi. state & $E_{\text{hull}}$ &  $T_{c}^{\text{AD}}$ & $\lambda$ & $\omega_{\text{log}}$ & $\text{TDOS}$ & $\text{PDOS}_\text{H}$ & $\text{PDOS}_\text{X}$ \\
\midrule
Y$_4$H$_{15}$  & +3 & 0.233  & 53.1  & 1.32 & 474  & 5.38 & 3.45 & 1.87 \\
Tb$_4$H$_{15}$ & +3 & 0.232  & 48.6  & 1.30 & 440  & 5.14 & 3.28 & 1.83 \\
Dy$_4$H$_{15}$ & +3 & 0.231  & 48.9  & 1.27 & 457  & 5.05 & 3.20 & 1.81 \\
Ho$_4$H$_{15}$ & +3 & 0.230  & 48.9  & 1.24 & 475  & 4.97 & 3.14 & 1.80 \\
Er$_4$H$_{15}$ & +3 & 0.228  & 49.3  & 1.23 & 482  & 4.90 & 3.08 & 1.79 \\
Tm$_4$H$_{15}$ & +3 & 0.227  & 50.4  & 1.23 & 495  & 4.83 & 3.02 & 1.78 \\
Lu$_4$H$_{15}$ & +3 & 0.225  & 51.3  & 1.19 & 523  & 4.70 & 2.92 & 1.75 \\
Th$_4$H$_{15}$ & +4 & 0.000  & 1.2   & 0.38 & 368  & 15.24 & 0.46 & 14.40 \\
Ti$_4$H$_{15}$ & +4 & 0.095  & 9.2   & 0.52 & 625  & 9.82 & 0.49 & 9.15 \\
Zr$_4$H$_{15}$ & +4 & 0.000  & 2.9   & 0.41 & 607  & 7.84 & 0.51 & 7.16 \\
Hf$_4$H$_{15}$ & +4 & 0.000  & 4.8   & 0.47 & 476  & 7.33 & 0.53 & 6.65 \\ 
Nb$_4$H$_{15}$ & +5 & 0.129  & 34.4  & 1.28 & 319  & 9.49 & 0.63 & 8.64 \\
\bottomrule
%Ta$_4$H$_{15}$ & +5 & 0.113  & 27.3  & 2.732 & 116  & 8.181 & 0.664 & 7.28 \\
\end{tabular*}
\end{table*}

\begin{figure}[htbp]
\begin{center}
\includegraphics[width=\columnwidth]{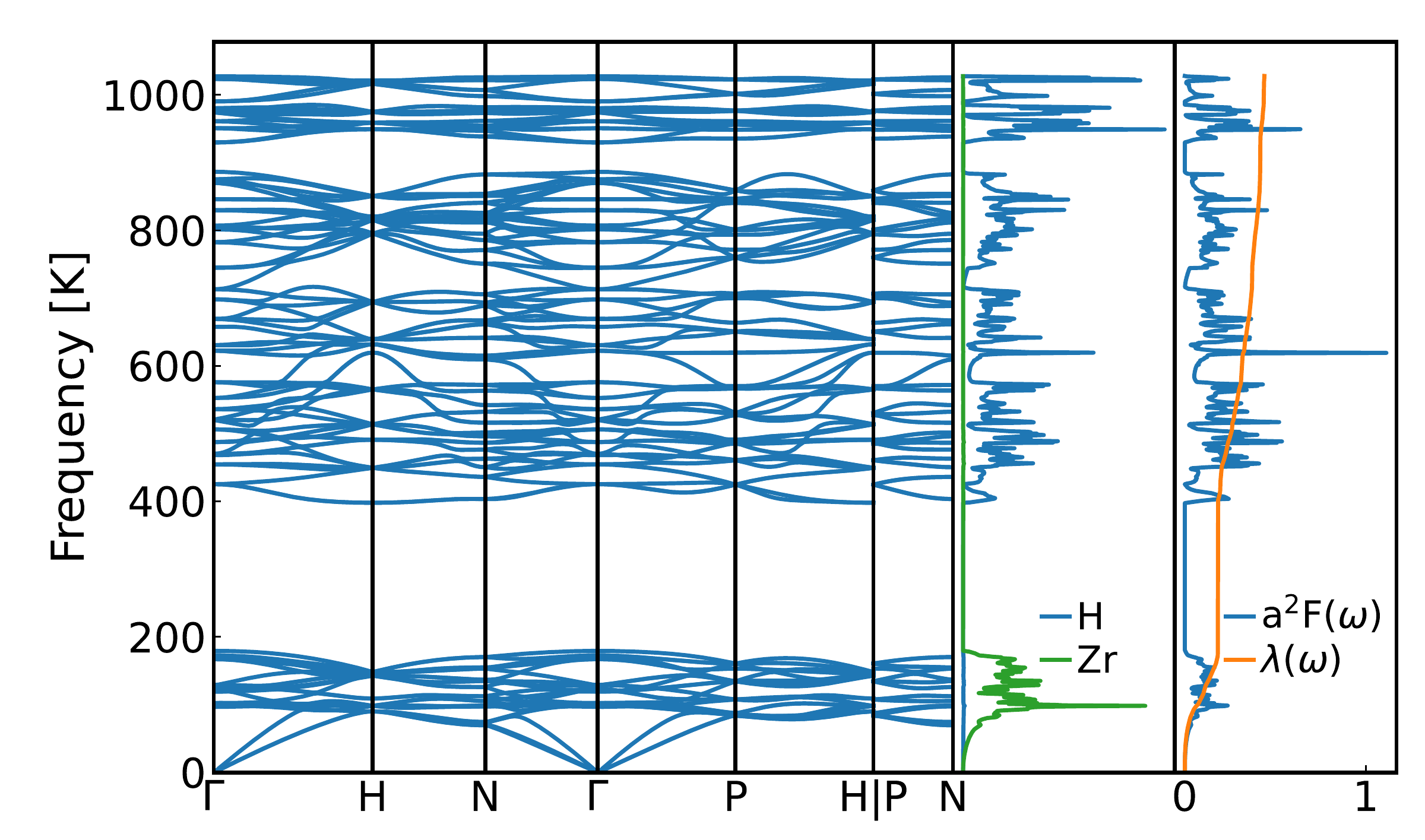}

(a)~\ce{Zr4H15}

\includegraphics[width=\columnwidth]{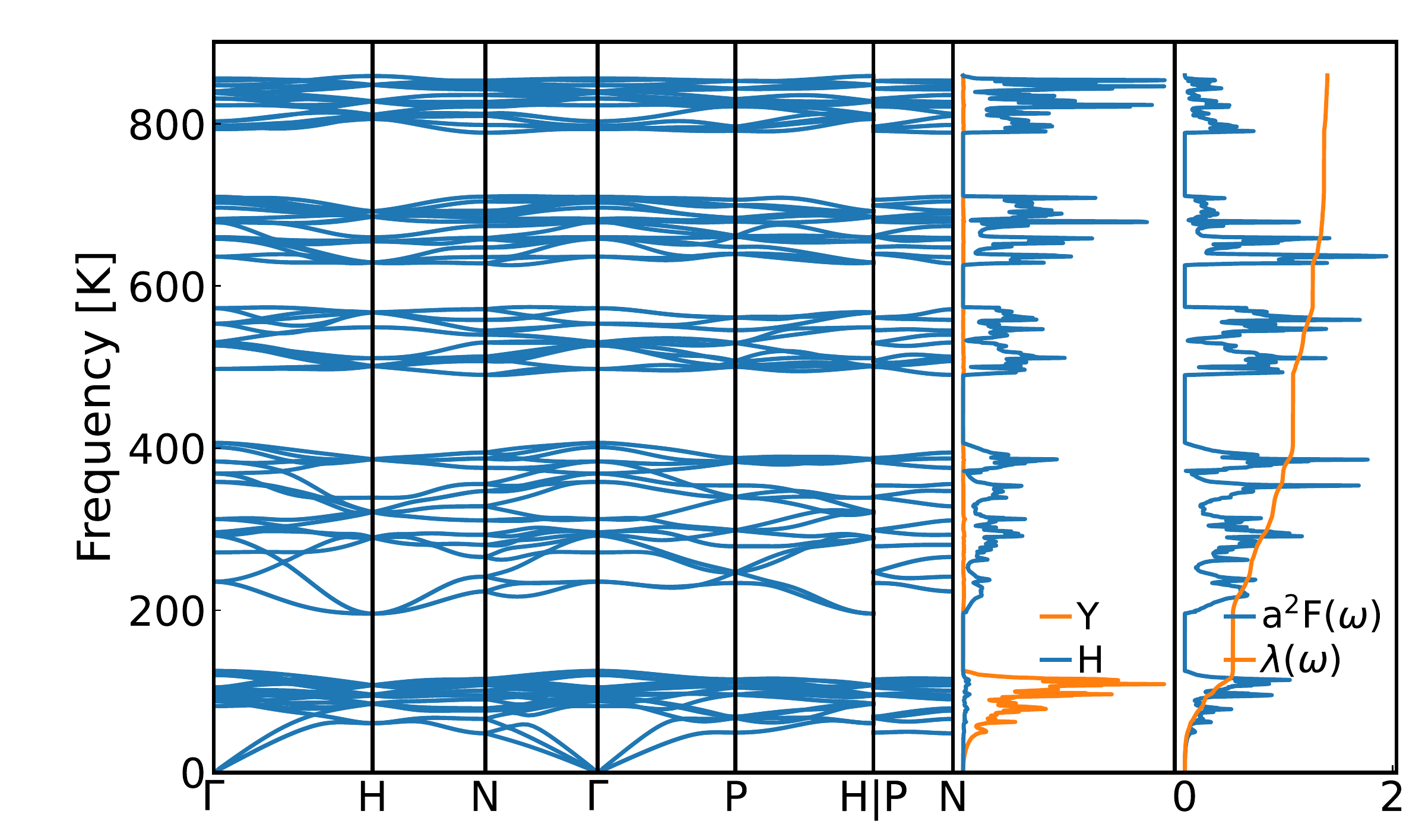}

(b)~\ce{Y4H15}
\end{center}
\caption{Phonon band structure, phonon density of states and Eliashberg spectral function $\alpha^2F(\omega)$ for representative compounds: (a) electron-doped \ce{Zr4H15} and (b) hole-doped \ce{Y4H15}. The stronger coupling with high-frequency hydrogen modes in the hole-doped compound is evident.}
\label{fig:ph-bs}
\end{figure}

We begin our discussion on superconductivity with compounds where X = Ti, Zr, Hf, and Th exists in the +4 oxidation state. This configuration yields one excess electron per formula unit occupying the conduction band. The electronic band structure of \ce{Zr4H15}, illustrated in \cref{fig:el-bs}, exemplifies a degenerate semiconductor, exhibiting striking similarities to that of the semiconducting compound \ce{YZr3H15}, presented in \cref{fig:el-bs}a (note that the different symmetry points and path in the Brillouin zone result from the reduced symmetry of the latter system). The phonon band structure, depicted in \cref{fig:ph-bs}, demonstrates characteristics typical of hydrides with heavy cations, where acoustic and low-lying optical modes consist exclusively of cation vibrations, while hydrogen governs the higher-energy modes. The maximum phonon frequency exceeds 2000~K. Nevertheless, despite the Fermi energy residing at a pronounced peak in the density of states, this compound demonstrates an unexpectedly small electron-phonon coupling constant ($\lambda$), resulting in exceedingly modest superconducting transition temperatures (see \cref{table:superconductors}). Analogous behavior manifests in other X$^{4+}$ materials. The calculated \Tc\ values for Zr and Hf compounds align closely with experimental measurements under pressure~\cite{kuzovnikov2019high,xie2020superconducting}. However, the predicted value for \ce{Th4H15} significantly underestimates the experimental observation (1.2~K versus the experimental 7--8~K~\cite{satterthwaite1970superconductivity, satterthwaite1972preparation}).

In contrast, \ce{Nb4H15} containing Nb$^{5+}$, introduces five additional electrons into the conduction band. This substantial electron doping profoundly transforms the electronic structure, although the fundamental band architecture of the hypothetical charge-compensated compound remains identifiable. The density of states at the Fermi level for this +5 compound is markedly elevated compared to its +4 counterparts, yielding electron-phonon coupling constants exceeding unity ($\lambda > 1$) and correspondingly enhanced $T_c$ values. Notably, the majority of coupling originates from the low-lying niobium phonon modes, rather than from the high-energy hydrogen vibrations, as demonstrated in \SIagm{agm073024595} in Supplementary Information (SI).

The most intriguing results emerge for compounds containing trivalent metals: \ce{Y4H15}, \ce{Tb4H15}, \ce{Dy4H15}, \ce{Ho4H15}, \ce{Er4H15}, \ce{Tm4H15}, and \ce{Lu4H15}. These systems, incorporating a +3 cation, generate three holes per formula unit in the valence band. All these compounds exhibit extraordinarily similar electronic and phononic band structures, culminating in nearly identical superconducting properties. The electronic band-structure of \ce{Y4H15} is illustrated in \cref{fig:el-bs}c. The resemblance to the charge compensated compound remains unmistakable, with the Fermi level positioned precisely on a prominent peak in the density of states. Once again, the low-energy phonon states (see \cref{fig:ph-bs}) originate from the heavy cation, while the hydrogen modes segregate into distinct manifolds. Due to the intricate geometry of these compounds, categorizing these vibrations proves challenging; nevertheless, spectral analysis reveals that the low-energy manifold predominantly exhibits torsional character, while the highest frequency modes correspond to hydrogen bond stretching vibrations. In this case, not only do the cation modes couple strongly to electrons, but the hydrogen modes contribute significantly as well, resulting in a substantial electron-phonon coupling constant ($\lambda=1.3$) and an exceptional superconducting transition temperature of approximately 50~K (see \cref{table:superconductors}).

The consistent behavior across diverse rare earth elements suggests that the superconducting properties are predominantly governed by common electronic structure features rather than the specific chemical identity of the metal. This fundamental observation receives further corroboration from our phonon calculations, which reveal strikingly similar phonon dispersions and electron-phonon coupling distributions across all trivalent metal compounds (see \cref{fig:ph-bs} and SI).

Our comprehensive analysis spanning multiple compounds demonstrates that hole doping proves more effective than electron doping for enhancing superconducting properties in X$_4$H$_{15}$ systems. The $\lambda$ values for hole-doped compounds consistently exceed those of electron-doped systems by a substantial margin, even when the total DOS at the Fermi level is comparatively lower (see \cref{table:superconductors}). The profound contrast between hole-doped and electron-doped systems becomes evident, with the former exhibiting $T_c$ values that surpass the latter by an order of magnitude.
\begin{figure}[h]
    \centering
    \includegraphics[width=0.9\columnwidth]{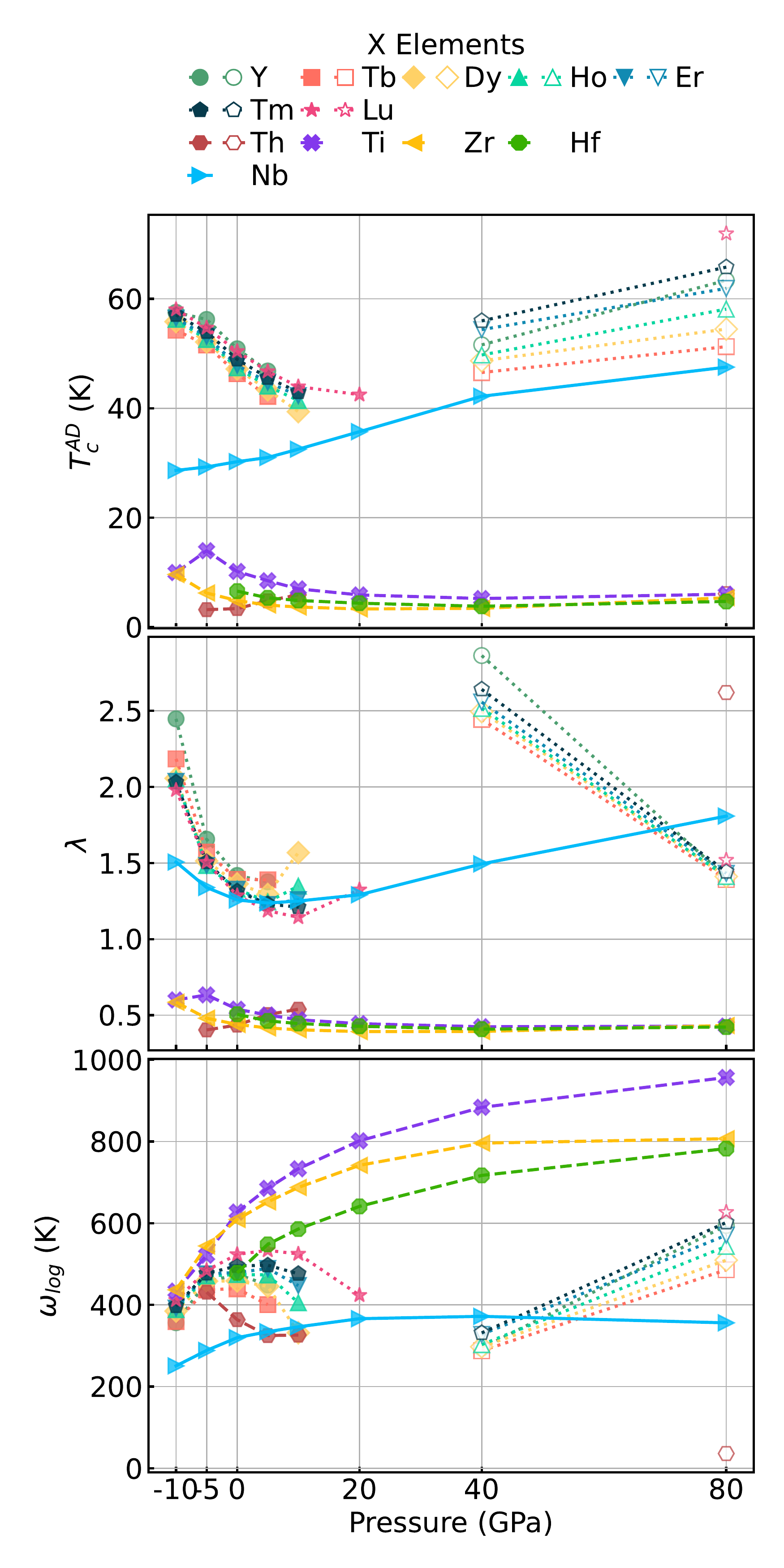}
    \caption{Superconducting critical temperature (\Tc), electron-phonon coupling constant (\la), and the logarithmic average of the phonon frequency (\olog) as a function of pressure. The filled and empty symbol represent the low- and high-pressure structures (See \cref{fig:structure}). For a certain pressure we omit the systems that are not dynamically stable. Note for Yb only PBE pseudopotential is available, for the sake of consistency we do not plot the curve of \ce{Yb4H15}. }
    \label{fig:tc-p}
\end{figure}

To understand how these materials behave under pressure, we performed calculations of the superconducting properties with varying pressure from -10 to 80~GPa (see \cref{fig:tc-p}). From the figure, the distinctive behavior of compounds depending on the oxidation state of the cation becomes immediately evident. For the +3 metals, \la\ decreases precipitously with pressure, while \olog\ remains relatively constant, resulting in a monotonic decrease of \Tc. This phenomenon can be attributed to the softening of a phonon mode that eventually becomes imaginary at pressures above 10--20~GPa. The compounds subsequently undergo a structural phase transition (see \cref{fig:structure}b) into a high-pressure phase that maintains the same $I\bar43d$ space group, but with cations reconfiguring into an undistorted bcc sublattice. In this structure, the electron-phonon coupling decreases with pressure, while \olog\ increases, leading to an enhancement of the transition temperature with pressure. For the +4 metals, \la\ diminishes to values substantially below 0.5, while \olog\ increases to values approaching 1000~K, resulting in a net reduction of \Tc\ with pressure. Finally, for Nb, \la\ exhibits a minimum at approximately 5~GPa, while \olog\ reaches a maximum at around 40~GPa, yielding a monotonically increasing \Tc\ that attains values exceeding 40~K at 80~GPa.

We must emphasize that while X$_4^{3+}$H$_{15}$ compounds demonstrate promising superconducting properties, they lack thermodynamic stability at ambient pressure. Our calculations reveal that these compounds reside approximately 230~meV/atom above the convex hull of stability, and this instability can be directly attributed to the charge imbalance and the resulting electronic structure, which favors decomposition into more stable phases. As anticipated, the most probable decomposition pathway involves the charge-compensated \ce{XH3} binary together with excess \ce{H2}.

While high-pressure synthesis represents one possible route for these compounds, we propose a potentially more direct approach. Our strategy involves initiating with \ce{YZr3H15}, which our calculations indicate lies remarkably close to thermodynamic stability and exhibits chemical plausibility due to its charge-compensated nature. Furthermore, cation site disorder will enhance stability through configurational entropy contributions. By synthesizing this compound with excess Y, one could effectively introduce hole doping while preserving structural integrity. As the DOS increases dramatically at the onset of valence states (see \cref{fig:el-bs}), even modest amounts of excess Y might suffice to induce hole-doped superconductivity in this system. Our band structure calculations demonstrate that removing merely 0.1--0.2 electrons per formula unit (equivalent to 5--10\% Y excess) could shift the Fermi level into the region of elevated DOS at the top of the valence band.

To rigorously evaluate this concept, we constructed ordered cells with \ce{Y2Zr2H15} stoichiometry. While these models cannot comprehensively capture the effects of substitutional doping in \ce{YZr3H15} due to the absence of disorder and the relatively high doping concentration, they provide critical insights into the potential behavior of such systems. The calculated electronic structure exhibits distinct metallic character with substantial DOS at the Fermi level, confirming the efficacy of this doping strategy. Furthermore, our phonon calculations for Y$_2$Zr$_2$H$_{15}$ reveal no imaginary modes, unequivocally demonstrating dynamic stability. The electron-phonon coupling constant for this compound is calculated to be approximately $\lambda=1.1$, corresponding to an estimated $T_c$ of around 40~K. This value, while lower than for pure \ce{Y4H15}, significantly exceeds that of any electron-doped compounds (as shown in SI).

In summary, our comprehensive computational investigation reveals that the superconducting properties of X$_4$H$_{15}$ compounds are profoundly influenced by their electronic configuration, with hole doping emerging as an exceptionally promising pathway to achieve high transition temperatures. The proposed strategy of introducing controlled hole doping in thermodynamically stable, charge-compensated compounds represents a practical approach for the experimental realization of these promising superconducting materials.

Furthermore, our findings demonstrate that the underlying structural stability of these compounds is intimately connected to their electronic configuration. Charge-compensated systems exhibit remarkable thermodynamic stability, while both electron and hole doping introduce varying degrees of instability that may necessitate high-pressure synthesis conditions.

The extraordinary consistency in superconducting properties across different trivalent metal compounds (Y, Dy, Er, Ho, Lu, Tb, Tm) strongly indicates that the hole-doping mechanism operates independently of the specific rare earth element. This universality reveals a fundamental electronic structure feature that could be systematically exploited in the rational design of new superconducting hydrides.

Our results carry significant implications for the broader field of hydride superconductivity. First, they underscore the critical importance of electronic structure engineering for optimizing superconducting properties. Second, they conclusively demonstrate that partial substitution constitutes an effective strategy for introducing carriers while maintaining structural stability. Finally, they suggest that other structurally related hydride systems could potentially benefit from analogous doping approaches.

\section{Methods}
\label{sec:methods}

We performed calculations of the entire family of \ce{X4H15} compounds where X spans the periodic table from Be to Bi, excluding rare-gases.
Geometry relaxations and total energy calculations were performed using the \textsc{vasp} code~\cite{vasp1,vasp2} with the Perdew-Burke-Ernzerhof approximation~\cite{Perdew2008} to the exchange-correlation functional. To sample the Brillouin zones, we implemented a 3$\times$3$\times$3 $\Gamma$-centred k-point grid. Spin-polarised calculations were initiated from a ferromagnetic configuration. We utilized the projector augmented wave (PAW) setup~\cite{paw1,paw2} within \textsc{vasp} version 5.2, applying a cutoff of 520~eV. We established the convergence criteria of the forces to be less than 0.005~eV/\AA.

Distances to the convex hull were calculated against the convex hull of the Alexandria database~\cite{alex1, alex2}. We remark that this represents the largest convex hull freely available, substantially more extensive than that of the Materials Project database~\cite{materialsproject}. All parameters, including pseudopotentials, were configured to ensure compatibility with the data available in the Alexandria database~\cite{alex1, alex2}. 

Phonon calculations were executed using version 7.1 of \textsc{Quantum Espresso}~\cite{Giannozzi2009,Giannozzi2017} with the Perdew-Burke-Ernzerhof functional for solids (PBEsol)~\cite{Perdew2008} generalized gradient approximation.
We employed the PBEsol pseudopotentials~\cite{Perdew2008} from the \textsc{pseudodojo} project~\cite{vanSetten2018pseudodojo}, specifically the stringent, scalar-relativistic norm-conserving set. 
Geometry optimizations were conducted using a uniform $\Gamma$-centered $4\times4\times4$ $k$-point grid. Convergence thresholds for energies, forces, and stresses were established at $1 \times 10^{-8}$~a.u., $1 \times 10^{-6}$~a.u., and $5 \times 10^{-2}$~kbar, respectively. For the electron-phonon coupling calculations, we implemented a double-grid technique, utilizing a $8\times8\times8$ $k$-grid as the coarse grid, and a $16\times16\times16$ as the fine grid. For the $q$-sampling of phonons, we employed a $2\times2\times2$ $q$-point grid. The double $\delta$-integration to obtain the Eliashberg function was performed with a Methfessel–Paxton smearing of 0.05~Ry.

\section{Acknowledgements}
K.G. acknowledges financial support from the China Scholarship Council. M.A.L.M. was supported by a grant from the Simons Foundation (SFI-MPS-NFS-00006741-12, P.T.) in the Simons Collaboration on New Frontiers in Superconductivity and by the Keele and the Klaus Tschira foundations as a part of the SuperC collaboration.
T.F.T.C. acknowledges the financial support from FCT - Fundação para a Ciência e Tecnologia, I.P. through the projects UIDB/04564/2020 and CEECINST00152/2018/CP1570/CT0006, with DOI identifiers 10.54499/UIDB/04564/2020 and 10.54499/CEECINST/00152/2018/CP1570/CT0006 respectively, and computing resources provided by the project Advanced Computing Project 2023.14294.CPCA.A3, platform Deucalion.
W.C., H.C.W and M.A.L.M. acknowledge the funding from the Sino-German Mobility Programme under Grant No. M-0362.
The authors thank the Gauss Centre for Supercomputing e.V. (www.gauss-centre.eu) for funding this project by providing computing time on the GCS supercomputer SUPERMUC-NG at the Leibniz Supercomputing Centre (www.lrz.de) under the project pn25co. H.C.W and M.A.L.M. would like to thank the NHR Centre PC2 for providing computing time on the Noctua 2 supercomputers. 
\newline

\bibliography{biblio}

\end{document}